\documentclass[11pt, a4paper]{article}
\usepackage{jheppub}
\usepackage[T1]{fontenc}
\usepackage{mathtools}
\usepackage{subcaption}
\usepackage{amsmath}
\usepackage{cleveref}
\usepackage[nolist]{acronym}
\usepackage{bm}
\usepackage{esvect}

\crefname{figure}{figure}{figures} % JHEP: figure should not be abbreviated. Do this instead of noabbrev package option to keep eq. abbreviations.

\graphicspath{{img/}}

\newcommand{\dd}{\text{d}}
\renewcommand{\i}{\ensuremath{\mathrm{i}}}
\newcommand{\eps}{\epsilon}
\newcommand{\incite}[1]{ref.~\cite{#1}}
\newcommand{\incites}[1]{refs.~\cite{#1}}

\newcommand{\bra}[1]{\langle #1|}
\newcommand{\ket}[1]{|#1 \rangle}
\newcommand{\braket}[1]{\langle #1 \rangle}

\newcommand{\braketsq}[1]{[ #1 ]}

\newcommand{\am}{\mathcal{A}}
\newcommand{\al}[1]{{\am^{(#1)}}^{\mu}}
\newcommand{\sal}[3]{{\am^{(#1)}_{#2, #3}}^{\mu}}
\newcommand{\nl}{n_l}

\title{Lepton-pair scattering with an off-shell and an on-shell photon at two loops in massless QED}

\author[a]{Simon Badger,}
\author[a,b]{Jakub Kry\'s,}
\author[a]{Ryan Moodie,}
\author[a]{Simone Zoia}

\affiliation[a]{
    Dipartimento di Fisica and Arnold-Regge Center, Università di Torino, and INFN, Sezione di Torino, Via P.\ Giuria 1, I-10125 Torino, Italy
}
\affiliation[b]{
    Institute for Particle Physics Phenomenology, Department of Physics, Durham University, Durham DH1 3LE, United Kingdom
}

\emailAdd{simondavid.badger@unito.it}
\emailAdd{jakubmarcin.krys@unito.it}
\emailAdd{ryaniain.moodie@unito.it}
\emailAdd{simone.zoia@unito.it}

\abstract{
    We compute the two-loop QED helicity amplitudes for the scattering of a
    lepton pair with an off-shell and an on-shell photon, $0\to\ell \bar\ell
    \gamma \gamma^*$, using the approximation of massless leptons. We express
    all master integrals relevant for the scattering of four massless particles
    with a single external off-shell leg up to two loops in a basis of
    algebraically independent \aclp{MPL}, which guarantees an efficient
    numerical evaluation and compact analytic representations of the
    amplitudes. Analytic forms of the amplitudes are reconstructed from
    numerical evaluations over finite fields. Our results complete the amplitude-level
    ingredients contributing to the \acs{N3LO} predictions of electron-muon
    scattering $e\mu\to e\mu$, which are required to meet the precision goal of
    the future MUonE experiment.
}

\begin{document}

\maketitle
\flushbottom

\begin{acronym}
    \acro{DE}{differential equation}
    \acro{HVP}{hadronic vacuum polarisation}
    \acro{IBP}{integration-by-parts}
    \acro{IR}{infrared}
    \acro{ISP}{irreducible scalar product}
    \acro{ISR}{initial state radiation}
    \acro{MI}{master integral}
    \acro{MPL}{multiple polylogarithm}
    \acro{N3LO}[N\textsuperscript{3}LO]{next-to-next-to-next-to-leading order}
    \acro{NLO}{next-to-leading order}
    \acro{NNLO}{next-to-next-to-leading order}
    \acro{QCD}{quantum chromodynamics}
    \acro{QED}{quantum electrodynamics}
    \acro{RVV}{real-double-virtual}
    \acro{SM}{Standard Model}
    \acro{UV}{ultraviolet}
    \acro{VV}{double-virtual}
\end{acronym}

\section{Introduction}
The MUonE experiment~\cite{Abbiendi:2016xup, Abbiendi:2022oks, Spedicato:2022qtw, Pilato:2022wvg} will measure the hadronic running of the electromagnetic coupling $\alpha$ using low-energy elastic electron-muon scattering, $e\mu \to e\mu$.
This will enable a new and precise determination of the \ac{HVP} contribution $a_{\mu}^{\text{HVP}}$~\cite{CarloniCalame:2015obs,Balzani:2021del} to the muon anomalous magnetic moment $a_{\mu}$. This is required in light of the recent tensions between experimental~\cite{Muong-2:2021ojo}, \ac{SM} data-driven~\cite{Aoyama:2020ynm}, and lattice \ac{QCD}~\cite{Borsanyi:2020mff} results for $a_\mu$.
Increasing the precision of the theoretical predictions for $e\mu \to e\mu$ scattering is a high priority for the planned MUonE experiment~\cite{Banerjee:2020tdt,Budassi:2022dco} and has seen good progress in the last few years~\cite{Alacevich:2018vez,Budassi:2021twh,Fael:2019nsf,Fael:2018dmz,CarloniCalame:2020yoz}.
The recent completion of full \ac{NNLO} \ac{QED} corrections~\cite{Broggio:2022htr} indicates that 
\ac{N3LO} corrections in differential distributions are required to meet MUonE's precision goal of 10 parts per million.
Electron-line corrections, meaning corrections to the subprocess with the muon line stripped off ($e\to e \gamma^*$), are the dominant corrections~\cite{Broggio:2022htr}, and a collaborative project was started to perform their fixed-order calculation at \ac{N3LO}~\cite{Durham:n3lo}.
With the triple-virtual corrections now available~\cite{Fael:2022rgm,Fael:2022miw,Fael:2023zqr}, the main missing ingredient is the \ac{RVV} matrix element ($e\to e \gamma \gamma^*$) at two loops.
While these contributions could be extracted from amplitudes in the literature~\cite{Garland:2001tf,Garland:2002ak,Gehrmann:2011ab}, our direct computation provides the massless \ac{RVV} contribution in a complete and compact form.

Another application of the $0\to \ell \bar\ell \gamma \gamma^*$ amplitudes is in electron-positron annihilation experiments~\cite{precisionsm}.
They are required for initial-state corrections in predictions of the ratio of hadron-to-muon production in $e^+e^-$ collisions, which is an important input for existing \ac{SM} predictions of $a^{\text{HVP}}_\mu$~\cite{Abbiendi:2022liz}.
The two-loop amplitudes contribute to \ac{RVV} corrections to $e^+e^+\to\gamma^*$ in direct scan measurements, while radiative return measurements concern corrections to $e^+e^-\to\gamma\gamma^*$~\cite{Aoyama:2020ynm}.
In the latter configuration, the $e^+e^-$ beam has a fixed centre-of-mass energy of a few GeV and the on-shell photon originates from \ac{ISR}.
The energy lost to the \ac{ISR} photon is used to effectively scan over the energies of the decay of the off-shell photon.
A differential cross section of, for example, $\gamma^*\to\text{hadrons}$ with respect to the centre-of-mass energy of the decay, $\dd\sigma/\dd s$, can be extracted from measurements of the differential cross section with respect to the energy of the \ac{ISR} photon, $\dd\sigma/\dd E_\gamma$.
State-of-the-art predictions for these measurements are currently at \ac{NLO}~\cite{Abbiendi:2022liz}.
We provide the two-loop $e^+e^-\to\gamma\gamma^*$ amplitudes required for the \ac{VV} corrections at \ac{NNLO}, although the bottleneck remains in the hadronic decay.

Our amplitudes are calculated in the approximation of massless leptons.
In the \ac{NNLO} massive $e\mu \to e\mu$ cross section calculation~\cite{Broggio:2022htr}, the authors obtain photonic corrections (those with no closed fermion loops), using a small-mass expansion~\cite{Penin:2005eh,Becher:2007cu,Engel:2018fsb} applied to the two-loop amplitudes with massless electrons for the \ac{VV} corrections.
This approximation relies on the electron mass being much smaller than any other scale, which is valid in the bulk of phase space.
Further splitting the photonic corrections, they take the subset of electron-line corrections and find that the relative difference to the true massive \ac{NNLO} differential cross section is generally around $10^{-3}\alpha^2$, where $\alpha$ is the fine-structure constant, which is negligible compared to the $10^{-5}$ precision goal.
The approximation breaks down in soft and collinear regions, where they treat the amplitudes using \ac{IR} factorisation~\cite{Banerjee:2021mty,Engel:2021ccn,Engel:2023ifn}, and is not used for contributions including closed fermion loops~\cite{Engel:2018fsb,Engel:2019nfw}.
Our amplitudes can be used analogously for the \ac{RVV} corrections at \ac{N3LO}.

\smallskip

Our computation uses the modern technology developed for \ac{QCD} amplitudes
with many scales. The high-multiplicity amplitude frontier in massless \ac{QCD}
lies with two-loop five-particle processes, with
leading-colour~\cite{Abreu:2019odu,Abreu:2020cwb,Chawdhry:2020for,Agarwal:2021grm,Abreu:2021oya,Chawdhry:2021mkw}
and
full-colour~\cite{Agarwal:2021vdh,Badger:2021imn,Badger:2023mgf,Abreu:2023bdp}
results in a form ready for phenomenological application becoming available
over the past few years. Recently, the first single-external-mass calculations
are also
appearing~\cite{Hartanto:2019uvl,Badger:2021nhg,Badger:2021ega,Badger:2022ncb}.
These computations have made extensive use of finite-field arithmetic to
sidestep large intermediate expressions. This technology has had a considerable
impact for solutions of systems of \ac{IBP}
identities~\cite{vonManteuffel:2014ixa,Klappert:2020nbg,Smirnov:2019qkx} but
also applies more widely to scattering amplitude
computations~\cite{Peraro:2016wsq,Peraro:2019svx}. Motivated by the improved
algorithms, we choose to implement a complete finite-field based reduction for
the $2\to 2$ processes with an off-shell leg. Since the kinematics are
relatively simple in comparison with other high-multiplicity configurations,
this technology is not essential. It does, however, provide an opportunity to
review the new techniques for readers who are not familiar with them.

A key ingredient for computing the scattering amplitudes are analytic
expressions for the required Feynman integrals.  Complete analytic results up
to two loops are already available in the
literature~\cite{gehrmann:2000zt,gehrmann:2001ck,Gehrmann:2002zr,Duhr:2012fh,Gehrmann:2023etk}.  Expansions
of these integrals up to higher orders in the dimensional regularisation
parameter $\epsilon$ have also been reconsidered
recently~\cite{Gehrmann:2023etk}, in view of their usage for \ac{N3LO}
corrections to $2\to 2$ processes in
\ac{QCD}~\cite{Gehrmann:2022vuk,Gehrmann:2023zpz}.  The state of the art for
integrals with this kinematic configuration has reached three
loops~\cite{DiVita:2014pza,Canko:2020gqp,Canko:2021xmn,Henn:2023vbd}.  We
revisit the computation of the one- and two-loop integrals following the
approach of
\incites{Gehrmann:2018yef,Chicherin:2020oor,Badger:2021nhg,Chicherin:2021dyp,Abreu:2023rco}
based on the construction of a \emph{basis} of independent special functions,
which gives a unique and uniform representation of all the required Feynman
integrals up to transcendental weight four.  This enables a more efficient
computation of the amplitudes using the modern workflow based on finite-field
arithmetic, and leads to more compact expressions. We give explicit expressions
for the basis functions in terms of \acp{MPL} which can be evaluated in an
efficient and stable way throughout the physical phase space.  We compute all
crossings of all massless one- and two-loop four-particle Feynman integrals
with an external off-shell leg, so that our results for the integrals may be of
use for any scattering process with these kinematics.

Our paper is organised as follows.
In \cref{sec:structure}, we describe our decomposition of the helicity amplitudes and detail how we express the off-shell currents.
In \cref{sec:calc}, we discuss our computation of analytic amplitudes by numerical evaluations over finite fields.
In \cref{sec:spec-fns}, we present the computation of the Feynman integrals in terms of a basis of special functions.
We draw our conclusions in \cref{sec:conc}.
We provide useful technical details in appendices. 
We define the relevant families of Feynman integrals in \cref{app:int_def}.
In \cref{app:altIBPs}, we discuss in detail how we handle permutations of the integral families in the \ac{IBP} reduction.
In \cref{app:mtvs}, we describe our rational parametrisation of the kinematics.
\Cref{app:poles} is devoted to the \ac{UV} renormalisation and \ac{IR} factorisation which determine the pole structure of the amplitudes.
In \cref{app:an_cont} we discuss the analytic continuation of the special functions to the physical kinematic regions.

\section{Structure of the amplitude}
\label{sec:structure}

We calculate the one- and two-loop \ac{QED} corrections to the process
\begin{align}
    \label{eq:scatter}
    0 \to \ell(p_1,h_1) + \bar{\ell}(p_2,h_2) + \gamma(p_3,h_3) + \gamma^{*}(p_4) \,,
\end{align}
which we call $0\to \ell \bar\ell \gamma \gamma^*$ for short. 
Here, $\ell$ denotes an on-shell massless lepton and $\gamma$ ($\gamma^*$) an on-shell (off-shell) photon, while $h_i$ and $p_i$ are the helicity and momentum of the $i\textsuperscript{th}$ particle.
We take the external momenta $p_i$ to be all outgoing. They satisfy the following momentum-conservation and on-shell conditions:
\begin{align}
    \sum_{i=1}^{4} p_i^\mu = 0 \,, \qquad \qquad \qquad p_i^2 = 0 \quad \forall \, i=1,2,3\,.
\end{align}
The single-off-shell four-particle phase space is described by three independent scalar invariants, which we choose as
\begin{align}
    \vec{s} \coloneqq \{s_{12}, s_{23}, s_4\} \,,
\end{align}
where $s_{i\ldots j} \coloneqq (p_i+\ldots+p_j)^2$.
We use dimensional regularisation in the 't~Hooft-Veltman scheme~\cite{Gnendiger:2017pys}, with $D=4-2\eps$ spacetime dimensions (where $\eps$ is the dimensional regulator) and four-dimensional external momenta.

Because of the off-shell photon in the process, the helicity amplitudes $\am^{\mu}(1_\ell,2_{\bar\ell},3_\gamma,4_{\gamma^*})$ are actually off-shell currents carrying a free Lorentz index.
We consider the perturbative \ac{QED} expansion of the helicity amplitudes,
\begin{align} \label{eq:loopdecomp}
    \am^{\mu}(1_\ell,2_{\bar\ell},3_\gamma,4_{\gamma^*}) = g_e^2 \sum_{L\ge 0} \left( n_\eps \frac{\alpha}{4\pi} \right)^L \al{L}(1_\ell,2_{\bar\ell},3_\gamma,4_{\gamma^*}) \,,
\end{align}
with prefactor $n_\eps=\i (4\pi)^\eps \mathrm{e}^{-\eps\gamma_E}$, electromagnetic coupling $g_e$, and $\alpha=g_e^2/(4\pi)$.
We truncate the expansion at $L=2$ loops.
We set the renormalisation scale $\mu_R$ to one throughout the computation and restore the dependence on it in the final analytic result by dimensional analysis. For the bare amplitudes we have that
\begin{align}
    \label{eq:scale-restoration}
    \al{L}(\mu_R) &= \mu_R^{2\eps L} \al{L}(\mu_R=1) \,.
\end{align}

There are two independent helicity configurations $(h_1,h_2,h_3)$, which we take as
\begin{align} \label{eq:helconfs}
    \{-+- \,, \ \ -++\} \,.
\end{align}
We derive the analytic expressions for these helicity amplitudes.
We obtain the remaining helicity configurations, $\{+-+ \,,\ \ +--\}$, through parity transformation (see appendix~C of \incite{Badger:2023mgf}).

\begin{figure}
    \begin{center}
        \begin{subfigure}[c]{0.3\linewidth}
            \centering
            \includegraphics[scale=0.8]{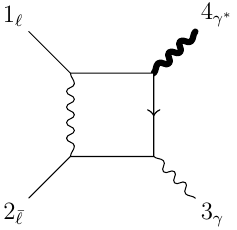}
            \caption{$\sal{1}{0}{0}$}
            \label{fig:1Lbox}
        \end{subfigure}
        \begin{subfigure}[c]{0.3\linewidth}
            \centering
            \vspace{4ex}
            \includegraphics[scale=0.8]{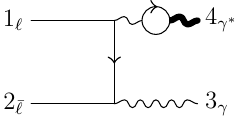}
            \vspace{4ex}
            \caption{$\sal{1}{1}{1}$}
            \label{fig:1Lbubble}
        \end{subfigure}
        \\
        \vspace{1em}
        \begin{subfigure}[c]{0.3\linewidth}
            \centering
            \includegraphics[scale=0.8]{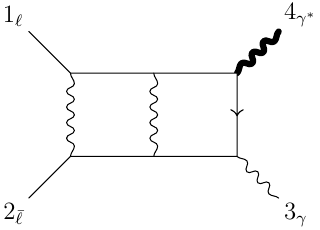}
            \caption{$\sal{2}{0}{0}$}
            \label{fig:2Lbox1}
        \end{subfigure}
        \begin{subfigure}[c]{0.3\linewidth}
            \centering
            \includegraphics[scale=0.8]{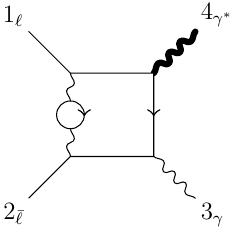}
            \caption{$\sal{2}{1}{0}$}
            \label{fig:2Lboxbubble1}
        \end{subfigure}
        \begin{subfigure}[c]{0.3\linewidth}
            \centering
            \includegraphics[scale=0.8]{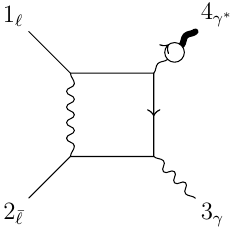}
            \caption{$\sal{2}{1}{1}$}
            \label{fig:2Lboxbubble2}
        \end{subfigure}
        \\
        \vspace{1em}
        \begin{subfigure}[c]{0.3\linewidth}
            \centering
            \includegraphics[scale=0.8]{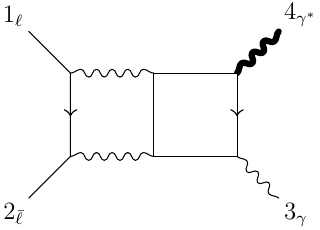}
            \caption{$\sal{2}{1}{2}$}
            \label{fig:2Lbox2}
        \end{subfigure}
        \begin{subfigure}[c]{0.3\linewidth}
            \centering
            \vspace{4ex}
            \includegraphics[scale=0.8]{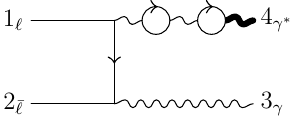}
            \vspace{4ex}
            \caption{$\sal{2}{2}{1}$}
        \end{subfigure}
        \caption{
            Representative Feynman diagrams for the subamplitudes defined in \cref{eq:nldecomp}.
            The off-shell external leg is indicated by a bold line.
        }
        \label{fig:repr-diagrams}
    \end{center}
\end{figure}

We decompose the loop-level helicity amplitudes $\al{L}$ into gauge-invariant
subamplitudes $\sal{L}{i}{j}$, where the subscript $i$ counts the number of
closed massless fermion loops and $j$ the number of external photons attached
to closed fermion loops. The non-zero contributions are
\begin{subequations}
    \label{eq:nldecomp}
    \begin{align}
        \al{1} &= \sal{1}{0}{0} + \nl \, \sal{1}{1}{1} \,, \\
        \al{2} &= \sal{2}{0}{0}
        + \nl \left( \sal{2}{1}{0} + \sal{2}{1}{1} + \sal{2}{1}{2} \right)
        + \nl^2 \sal{2}{2}{1} \,,
    \end{align}
\end{subequations}
where $\nl$ denotes the number of charged lepton flavours running in the loops.
Representative Feynman diagrams contributing to these subamplitudes are illustrated in \cref{fig:repr-diagrams}.
Amplitudes with a closed fermion loop attached to an odd number of photons vanish by Furry's theorem.

We decompose the amplitude and subamplitude currents as
\begin{align}
    \label{eq:offshellproj}
    \al{L} = \sum_{k=1}^4 a_{k}^{(L)} \, q_k^\mu \,, \qquad \qquad \sal{L}{i}{j} = \sum_{k=1}^4 a_{i,j;k}^{(L)} \, q_k^\mu \,,
\end{align}
using the following basis written with the spinor-helicity formalism:
\begin{align}
    \label{eq:proj-basis}
    q_k^\mu &= p_k^\mu \quad\forall \, k=1,2,3 \,, &
    q_4^\mu &= \frac{\langle 2|p_3p_1\sigma^\mu|2 ] - \langle 1|p_3p_2\sigma^\mu|1 ]}{2 s_{12}} \,.
\end{align}
Readers not familiar with the spinor-helicity formalism may like to consult one of the many good reviews on the subject~\cite{Mangano:1990by,Dixon:1996wi,Badger:2023eqz}.
Note that $q_4$ is orthogonal to the momenta $p_i$ by construction; one can in fact show that $q_4^\mu\propto \varepsilon^{\mu\nu\rho\sigma}{q_1}_\nu{q_2}_\rho{q_3}_\sigma$. 
The subamplitude coefficients $a_{i,j;k}^{(L)}$ can be related to the amplitude ones $a_{k}^{(L)}$ through \cref{eq:nldecomp}.

\smallskip

The scattering amplitudes $\mathcal{M}^{(L)}$ for fully on-shell processes (for instance, for $0 \to e^-e^+ \gamma \mu^-\mu^+$) are obtained by contracting the amplitude currents $\al{L}$ (for $0 \to e^-e^+ \gamma \gamma^*$) with a suitable decay current $\mathcal{V}_\mu$ (in this example, $\gamma^*\to\mu^-\mu^+$), as
\begin{align}
    \label{eq:on-shell-amps}
    \mathcal{M}^{(L)} &\coloneqq \mathcal{A}^{(L)} \cdot\mathcal{V}  = \sum_{k=1}^4 a_k^{(L)} \, \left( q_k\cdot\mathcal{V} \right) \,.
\end{align}
In this manner, the on-shell amplitudes $\mathcal{M}^{(L)}$ are given by the scalar product between the vector of coefficients $(a_1^{(L)}, \ldots, a_4^{(L)})$, and that of decay-vector contractions $(q_1\cdot\mathcal{V}, \ldots, q_4\cdot\mathcal{V} )$.
The coefficients $a^{(L)}_k$ depend on the helicities of the three on-shell particles in \cref{eq:scatter}, while the decay vector $\mathcal{V}_{\mu}$ depends on the helicities of the particles the off-shell photon decays to.
The helicity-summed interference between the $L_1$-loop and the $L_2$-loop matrix elements is then given by
\begin{align} \label{eq:squared_M}
    \mathcal{M}^{(L_1,L_2)} &= \frac{1}{4} \sum_{\vec{h}} {\mathcal{M}^{(L_1)}_{\vec{h}}}^* \mathcal{M}^{(L_2)}_{\vec{h}} \,,
\end{align}
where the subscripts $\vec{h}$ indicates the helicities of all on-shell particles --- that is, including the decay products of the off-shell photon --- and the overall constant factor averages over the helicities of the incoming particles.

\smallskip

The output of the computation described in \cref{sec:calc} is the set of four projections $\mathcal{A}^{(L)}_{i,j} \cdot q_k$ for each helicity configuration listed in \cref{eq:helconfs}.
From these, we determine the subamplitude coefficients $a_{i,j;k}^{(L)}$ by inverting \cref{eq:offshellproj}, as
\begin{align}
    \label{eq:ampcoeffinv}
    a_{i,j ; k}^{(L)} &= \sum_{m=1}^4 \left(\mathrm{G}^{-1}\right)_{km} \left(\mathcal{A}^{(L)}_{i,j} \cdot q_m\right) \,,
\end{align}
where $\mathrm{G}$ is the Gram matrix of the vectors $q_i$, that is, the matrix of entries $\mathrm{G}_{ij} \coloneqq q_i \cdot q_j$ for $i,j=1,\ldots,4$.
At loop level, we write the subamplitude coefficients as
\begin{align}
    \label{eq:ampprojcoeffs}
    a_{i,j;k}^{(L)} &= \sum_{w=-2L}^{4-2L} \sum_r\,c_{r,w}\,\text{mon}_r(F)\,\eps^w ,
\end{align}
where $\text{mon}_r(F)$ are monomials of special functions $F$ (see \cref{sec:spec-fns}), and the coefficients $c_{r,w}$ are rational functions of the kinematics. 
We drop the dependence on $i$, $j$, $k$, and $L$ on the right-hand side of \cref{eq:ampprojcoeffs} for compactness. 
We truncate the Laurent expansion around $\eps=0$ to the orders required for computing \ac{NNLO} predictions.
We express the coefficients $c_{r,w}$ as $\mathbb{Q}$-linear combinations of a smaller set of linearly-independent coefficients (see \cref{sec:calc}).
The analytic expressions of the latter are given explicitly in terms of momentum twistor variables (see \cref{app:mtvs}).
We simplify these expressions through a multivariate partial fraction decomposition using \texttt{MultivariateApart}~\cite{Heller:2021qkz}, and by collecting the common factors.

In the ancillary files~\cite{zenodo}, the directory \path{amplitudes/} contains \texttt{Mathematica} files describing the bare helicity subamplitude currents $\sal{L}{i}{j}$ by their coefficients $a_{i,j;k}^{(L)}$ in the form of \cref{eq:ampprojcoeffs}.
The \texttt{Mathematica} script \path{current.m} is a reference implementation of the numerical evaluation of the bare amplitude coefficients $a_k^{(L)}$ in \cref{eq:ampprojcoeffs}, including summation of subamplitudes in \cref{eq:nldecomp}, treatment of dependent helicities, and renormalisation scale restoration in \cref{eq:scale-restoration}.
The \texttt{Mathematica} script \path{evaluation.wl} demonstrates the construction of the five-particle on-shell amplitudes in \cref{eq:on-shell-amps} for the process $0\to e^- e^+ \gamma \mu^- \mu^+$, and their helicity-summation to obtain the squared matrix elements in \cref{eq:squared_M}.
The results of the script are checked against a reference point included in \path{reference_point.json}.

\smallskip

We perform the following checks of our amplitudes.
\begin{description}
    \item[Ward identity]
        We verify the gauge invariance of the subamplitudes $\sal{L}{i}{j}$ by checking that they vanish on replacing the on-shell photon's polarisation vector with its momentum.

    \item[One-loop crosscheck]
        We successfully crosscheck our one-loop $\nl=0$ helicity-summed matrix element contracted with the decay $\gamma^*\to\mu^-\mu^+$ against the \ac{QED} \ac{NLO} electron-line corrections for $e\mu\to e\mu\gamma$ obtained with \texttt{McMule}~\cite{Banerjee:2020rww,ulrich_yannick_2022_6046769}.

    \item[Finite remainder]
      We verify that the $\eps$-poles of the bare amplitudes have the structure predicted by \ac{UV} renormalisation and \ac{IR} factorisation~\cite{Catani:1998bh,Gardi:2009qi,Gardi:2009zv,Becher:2009cu,Becher:2009qa}. We then subtract the expected poles and define finite remainders at one and two loops as
        \begin{subequations}
            \label{eq:finite-remainders}
            \begin{align}
                {\mathcal{F}^{(1)}}^\mu &= \left[ \al{1} - \frac{3}{2}\frac{\beta_0}{\eps} \al{0} \right] - Z^{(1)} \al{0} \,,\\
                {\mathcal{F}^{(2)}}^\mu &= \left[ \al{2} - \frac{5}{2}\frac{\beta_0}{\eps} \al{1} - \left(-\frac{15}{8}\frac{\beta_0^2}{\eps^2}+\frac{3}{4}\frac{\beta_1}{\eps}\right) \al{0} \right] - Z^{(2)} \al{0} - Z^{(1)} {\mathcal{F}^{(1)}}^\mu \,,
            \end{align}
        \end{subequations}
        where the square brackets separate renormalisation of \ac{UV} poles from subtraction of \ac{IR} poles.
        We present the derivation of these formulae in \cref{app:poles}.
\end{description}

\section{Setup of the calculation}
\label{sec:calc}

In this section, we outline the workflow we used to calculate our amplitudes.
Firstly, we generate all Feynman diagrams contributing to \cref{eq:scatter}
using \texttt{QGRAF}~\cite{Nogueira:1991ex}. Each diagram is then replaced with
the corresponding Feynman rules for vertices, propagators, and external states,
leading to a collection of $D$-dimensional Feynman integrals. Next, we filter
the integrals according to \cref{eq:loopdecomp,eq:nldecomp} using a collection
of \texttt{Mathematica} and \texttt{FORM} scripts~\cite{Kuipers:2012rf,
Ruijl:2017dtg}. Within each subamplitude $\sal{L}{i}{j}$, we then collect the
integrals according to their topology, by which we mean a unique set of
denominators. For example, the diagrams in
\cref{fig:2Lboxbubble1,fig:2Lboxbubble2} belong to different topologies, but
those in \cref{fig:2Lbox1,fig:2Lbox2} belong to the same topology (under the
assumption of massless lepton propagators). At this point, the subamplitudes
are sums of Feynman integrals over distinct integral topologies, with the
numerators given by linear combinations of monomials that depend on the loop as
well as the external momenta. To work with the projected helicity subamplitudes
$\am^{(L)}_{i,j}\cdot q_k$, we specify the polarisations of external particles
according to \cref{eq:helconfs}, as well as the projector $q_k^\mu$ of the
off-shell photon from \cref{eq:proj-basis}. It is natural to express
helicity-dependent objects using the spinor-helicity formalism. Then, the
monomials of loop momenta contain the following scalar products and spinor
strings:
\begin{equation} \label{eq:loopmomobjects}
	\left\{ k_i \cdot k_j, \,
    k_i \cdot p_j ,\,
    \braket{ij},\,
    \braketsq{ij},\,
    \langle i |k_i |j],\,
    \langle i |p_4 |j],\,
    \bra{i}k_i p_4\ket{j},\,
    [i| k_i p_4 |j] \right\} \,.
\end{equation}
Their coefficients, on the other hand, are composed of the same type of
objects, but do not contain any dependence on loop momenta $k_i$.  We express
these coefficients using the rational parametrisation of the kinematics
discussed in \cref{app:mtvs}. 

This marks the start of our finite-field sampling
procedure~\cite{Peraro:2016wsq}.  The goal of this approach is to sidestep the
algebraic complexity which typically plagues the intermediate stages of
symbolic computations by evaluating numerically all rational coefficients.
Using integers modulo some large prime number ---~which constitute a finite
field~--- for the numerical evaluation allows us to avoid the loss of accuracy
inherent to floating-point numbers, as well as the expensive
arbitrary-precision arithmetic required by rational numbers.  Manipulations
needed to further process the rational coefficients are a completely separate
problem from the calculation of the integrals or special functions that these
coefficients multiply. In fact, they can be implemented as a series of rational
transformations over finite fields. We stress that this is the methodology we
follow at each step of the computation described below. In particular, we use
the package \texttt{FiniteFlow}~\cite{Peraro:2019svx}, which is conveniently
interfaced to \texttt{Mathematica}. The analytic form of the coefficients is
not known at any intermediate step. It is reconstructed from the finite-field
samples only at the very end of the workflow.

\begin{figure}
    \begin{center}
        \begin{subfigure}[c]{\linewidth}
            \centering
            \includegraphics[scale=0.8]{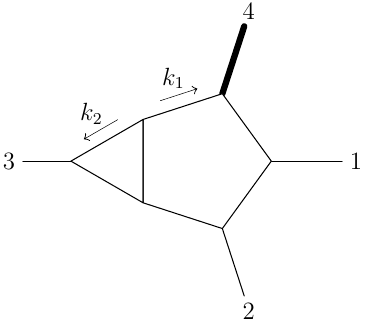}
            \includegraphics[scale=0.8]{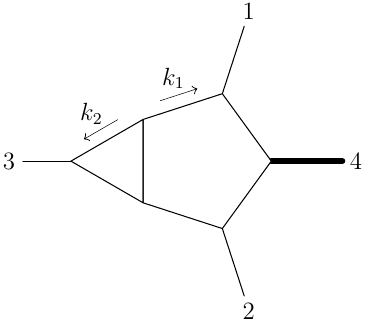}
            \includegraphics[scale=0.8]{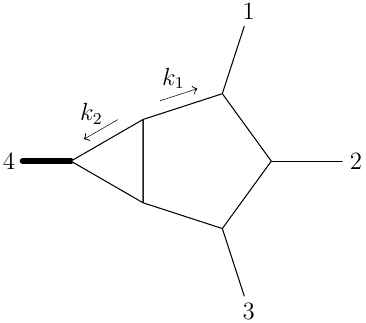}
            \caption{Penta-triangles}
            \label{fig:pentatriangle}
        \end{subfigure}
        \\
        \vspace{1em}
        \begin{subfigure}[c]{0.3\linewidth}
            \centering
            \includegraphics[scale=0.8]{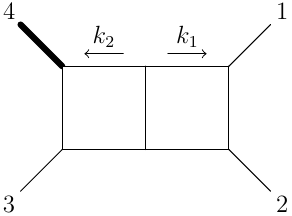}
            \caption{Double-box}
            \label{fig:double-box}
        \end{subfigure}
        \begin{subfigure}[c]{0.6\linewidth}
            \centering
            \includegraphics[scale=0.8]{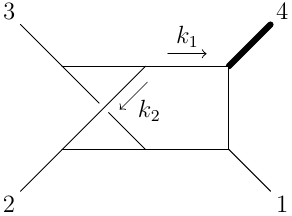}
            \includegraphics[scale=0.8]{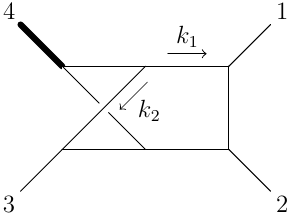}
            \caption{Crossed double-boxes}
            \label{fig:crosseddouble-box}
        \end{subfigure}
    \end{center}
    \caption{ 
        The two-loop amplitudes include six ordered integral families: three penta-triangles, a double-box, and two crossed double-boxes.
        All are planar except for the crossed double-boxes.
        The off-shell external leg is indicated by a bold line.
        External legs have outgoing momenta.
    }
    \label{fig:int-fams}
\end{figure}

Firstly, we note that not all integral topologies are independent: some of them
can be written as subtopologies of others. For this reason, we define the set
of \textit{maximal} topologies, i.e.~topologies with the maximum number of
propagators allowed for $L$-loop, $n$-particle diagrams.  In figure
\ref{fig:int-fams}, we present the maximal topologies for the process under
consideration in an arbitrary ordering of the external momenta (we give their
explicit definitions in \cref{app:int_def}). Several orderings of the external
momenta are relevant for the amplitudes, and we treat them as distinct
families. Next, we map all topologies present so far onto one of these maximal
topologies. The loop momenta dependent objects of \cref{eq:loopmomobjects} are
then expressed through the nine inverse propagators and \acp{ISP} associated
with the chosen maximal topology. In this way, each subamplitude is now a sum
of integrals compatible with \ac{IBP}
reduction~\cite{Tkachov:1981wb,Chetyrkin:1981qh}, while their coefficients
depend on the external kinematics and $\eps$. We generate the required \ac{IBP}
relations using \texttt{LiteRed}~\cite{Lee:2012cn}. The resulting \ac{IBP}
system is then solved using the Laporta algorithm~\cite{Laporta:2001dd} with
\texttt{FiniteFlow}'s linear solver to yield the reduction of all the integrals
present within our maximal topologies onto a much smaller subset of \acp{MI}.
It is beneficial to choose the \acp{MI} such that they satisfy \acp{DE} in canonical
form~\cite{Henn:2013pwa} (see \cref{sec:basis_construction}). This property
allows us to make the most of the optimisation measures we discuss below.
We stress that
the \ac{IBP} reduction is also done numerically over finite fields, since the
coefficients of the \ac{IBP} relations are rational functions of external
kinematics and the dimensional regulator $\eps$. This is an important
simplification, since analytic \ac{IBP} reduction often proves to be the
bottleneck of amplitude computations. For many amplitude applications, multiple permutations of the ordered topologies can appear. We outline a strategy to optimise the reduction in such situations in appendix \ref{app:altIBPs}.

At this point, each projected helicity subamplitude $\am^{(L)}_{i,j}\cdot q_k$
is written as a linear combination of \acp{MI} multiplied by rational
coefficients of $\eps$ and the kinematic variables. We now write the \acp{MI}
in terms of a basis of special functions up to the required order in $\eps$
(see \cref{sec:spec-fns}). Finally, we Laurent expand the amplitude around
$\eps=0$, the deepest pole being $1/\eps^{2 L}$ at $L$ loops. The only task
left is to reconstruct the rational coefficients of the special-function
monomials from their samples over finite fields. In general, this might be a
daunting challenge and its complexity stems from two separate factors. The
workflow described so far is a series of rational operations chained together
within a so-called dataflow graph~\cite{Peraro:2019svx}. As such, we
essentially have a black-box algorithm which takes numerical values of the
kinematic variables as input, and returns the corresponding numerical values of
the rational coefficients of the special-function monomials. The first factor
is that several sample points are necessary to infer the analytic expression of
these coefficients from their values in the finite fields. The required number
is correlated with the polynomial degrees of the rational functions viewed as
ratios of polynomials: the higher the degree, the more sample points are
required. The second factor affecting the reconstruction complexity is the time
it takes to obtain the values of the coefficients at each sample point. The
more complicated the dataflow graph is, i.e.~the more operations are chained
together and the more difficult each operation is, the longer it will take to
run the black-box algorithm. The most expensive operation in this regard is the
evaluation of the solution to the \ac{IBP} system.  The total reconstruction
time can thus be estimated as:
\begin{equation} \label{eq:rectimeschematic}
	\text{reconstruction time} \approx (\text{number of sample points}) \times (\text{evaluation time per point})\,.
\end{equation}
We emphasise that the sample evaluations can be run in parallel.  For a
detailed discussion of various strategies to improve the reconstruction time,
see section~4 of \incite{Badger:2021imn} and section~3.5 of
\incite{Badger:2022ncb}. Here, we give a brief overview of the tools that
proved sufficient for this work.

First, we look for $\mathbb{Q}$-linear relations among the rational
coefficients of each helicity subamplitude. This typically requires few sample
points with respect to the full reconstruction. We then solve these linear
relations to express all coefficients in terms of a minimal subset of
independent ones. Only the latter need to be reconstructed. Choosing them so
that they have the lowest degrees often leads to a decrease in the complexity
of the reconstruction.

The second strategy we employ is to match the rational coefficients with factorised ans\"atze informed by the singularity structure of Feynman integrals. The singularities of Feynman integrals can in fact be read off from the \acp{DE} they satisfy. For each coefficient we then write an ansatz made of the following factors:
\begin{equation}
	\left\{\braket{ij} , \, \braketsq{ij} , \, \bra{i}p_4|j] , \, s_{ij} - s_{k4} , \, s_{i4} - s_4 , \, s_4 \right\} \,,
\end{equation}
for all $i, j, k = 1, 2, 3$ such that $i\neq j \neq k$. This list includes denominator factors of the \acp{DE} satisfied by the \acp{MI} (listed by \cref{eq:alphabet}), 
as well as spinor structures aimed at capturing the phase information of helicity amplitudes.
We then determine the exponents of the ans\"atze by comparing them to the coefficients reconstructed on a univariate slice of the kinematic variables~\cite{Abreu:2019odu}, which are very cheap to obtain. We find that with this ansatz it is possible to determine all denominator factors ---~which indeed are linked to the singularity structure of the amplitude~--- and sometimes also some numerator factors. As a result, the undetermined functions yet to be reconstructed are of lower degree and require fewer sample points. 
We reconstruct the analytic form of the remaining rational functions using \texttt{FiniteFlow}'s built-in multivariate functional reconstruction algorithm.

Finally, we note that, for more computationally demanding processes, further ansatz-based techniques ---~for instance based on 
partial fraction decompositions or informed by the singularity structure of the amplitudes~--- may be used to optimise 
the functional reconstruction; see, for example, \incites{Badger:2021imn, Badger:2021ega, Badger:2022ncb,DeLaurentis:2022otd,Badger:2023mgf,Abreu:2023bdp,Liu:2023cgs}.

\section{Computation of the \aclp{MI}}
\label{sec:spec-fns}

The \acp{MI} for the relevant integral families were first computed analytically in
\incites{gehrmann:2000zt,gehrmann:2001ck}. The basis of special functions for
four-particle kinematics with an off-shell leg has been subsequently studied in
great detail with attention to the compactness of representations in terms of
polylogarithms and numerical evaluation across a complete phase-space~\cite{Duhr:2012fh,Gehrmann:2013vga,Gehrmann:2023etk} (see
also~\incite{Gehrmann:2002zr} for a thorough discussion of the analytic
continuation). We revisit this computation to obtain expressions for the
\acp{MI} which are well suited for the amplitude-computation workflow discussed
in \cref{sec:calc}. To this end, we compute the \acp{MI} for \emph{all
permutations} of the external legs in terms of a \emph{basis} of special
functions, following the approach of
\incites{Gehrmann:2018yef,Chicherin:2020oor,Badger:2021nhg,Chicherin:2021dyp,Abreu:2023rco}.
In other words, we express all the Feynman integrals contributing to the amplitudes in terms of a set of special functions which are (algebraically) independent.
Having such a unified and unique representation for all permutations of the integral families allows for simplifications and cancellations among different permutations of the Feynman integrals.
This leads to a simpler expression of the amplitudes and to a more efficient functional reconstruction in the finite-field setup presented in \cref{sec:calc}. 
We emphasise that our results cover all \acp{MI} required for computing \emph{any} two-loop four-particle amplitude with a single external off-shell leg, and not just the ones required for the amplitudes presented in this work.

We discuss the construction of the basis in \cref{sec:basis_construction}, and how we express it in terms of \acfp{MPL} in \cref{sec:basis_MPLs}. 
Finally, we give some details about the numerical evaluation and the checks we performed in \cref{sec:performance}.

\subsection{Construction of the special function basis}
\label{sec:basis_construction}

We follow the strategy presented in \incite{Abreu:2023rco}. The starting point are the \acp{DE} satisfied by the \acp{MI} for each family~\cite{Barucchi:1973zm,KOTIKOV1991158,KOTIKOV1991123,Gehrmann:1999as,Bern:1993kr}. Let $\tau$ label an integral family, e.g.\ the double-box in \cref{fig:double-box} for an arbitrary permutation of the external massless momenta. We choose a basis of \emph{pure} \acp{MI} $\vec{g}_{\tau}$, that is, a basis which satisfies \acp{DE} in the canonical form~\cite{Henn:2013pwa}
\begin{align}
\label{eq:canonicalDEs}
\dd \, \vec{g}_{\tau}(\vec{s}; \eps) = \eps \, \left( \sum_{i=1}^{7} \mathrm{A}^{(\tau)}_i \, \dd \log W_i(\vec{s}) \right) \cdot \vec{g}_{\tau}(\vec{s}; \eps) \,.
\end{align}
Here, $\dd$ is the total differential, $\dd f \coloneqq \dd s_{12} \, \partial_{s_{12}} f + \dd s_{23} \, \partial_{s_{23}} f +  \dd s_{4} \, \partial_{s_{4}} f $, $\mathrm{A}^{(\tau)}_i$ are constant $n_{\tau} \times n_{\tau}$ matrices, with $n_{\tau}$ the number of \acp{MI} of the family $\tau$, and 
\begin{align}
\label{eq:alphabet}
\begin{alignedat}{4}
& W_1 = s_{12} \,, 
&& W_2 = s_{23} \,, 
&& W_3 = s_{12} + s_{23} \,, \qquad
&& W_4 = s_{12} - s_4 \,,  \\
& W_5 = s_{23} - s_4 \,, \qquad
&& W_6 = s_{12} + s_{23} - s_4 \,, \qquad
&& W_7 = s_4 \, &&
\end{alignedat}
\end{align}
are called \emph{letters}. We emphasise that this alphabet covers all permutations of the relevant integrals. Specific integral families may contain only
subsets of it.
Canonical bases for the relevant integral families are already available in the literature~\cite{Canko:2021xmn,Gehrmann:2023etk,Henn:2023vbd}.
Given that, by today's standards, finding canonical bases for these integral families is simple, we re-derived them using a mixture of methods: the package \texttt{DlogBasis}~\cite{Henn:2020lye}, the analysis of results in the literature for related integral families (massless two-loop five-point planar integrals~\cite{Gehrmann:2015bfy} and two-loop four-point integrals with two massive external legs~\cite{Henn:2014lfa,Caola:2014lpa}), and a set of heuristic rules (see e.g.\ \incite{Dlapa:2022nct}).
We normalise the \acp{MI} such that their expansion around $\eps=0$ starts from order $\eps^0$,
\begin{align}
\vec{g}_{\tau}(\vec{s} ; \eps) = \sum_{w \ge 0} \eps^w \, \vec{g}^{(w)}_{\tau}(\vec{s}) \,.
\end{align}
For the purpose of computing two-loop scattering amplitudes up to their finite part (i.e., up to order $\eps^0$), it suffices to restrict our attention to $w \le 4$.
Since the \acp{MI} satisfy canonical \acp{DE}~\labelcref{eq:canonicalDEs}, the $\eps$-order of the \ac{MI} coefficients $\vec{g}^{(w)}_{\tau}(\vec{s})$ equals their \emph{transcendental weight}~\cite{Henn:2013pwa}.
We compute the derivatives of the \acp{MI} using \texttt{FiniteFlow}~\cite{Peraro:2019svx} and \texttt{LiteRed}~\cite{Lee:2012cn}.
We do so only for the integral families with the ordering of the external momenta shown in \cref{fig:int-fams}, and obtain those for all other orderings of the external massless legs by permutation. 
We provide the definition of the pure \acp{MI} and the corresponding \acp{DE} for all one- and two-loop four-point one-mass families in \cref{fig:int-fams} in the folder \path{pure_mi_bases/} of our ancillary files~\cite{zenodo}. 

In order to solve the \acp{DE}~\labelcref{eq:canonicalDEs} we need boundary values, i.e., values of all \acp{MI} up to order $\eps^4$ at a phase-space point. 
Due to the simplicity ---~by today's standards~--- of the integrals under consideration, an arbitrary (non-singular) phase-space point would do. 
Nonetheless, we make a more refined choice following some of the criteria of \incites{Chicherin:2020oor,Chicherin:2021dyp}.
We choose the following point in the $s_{12}$ channel (see \cref{app:an_cont}),
\begin{align} \label{eq:s0}
\vec{s}_0 = \left( 2, \, -\frac{1}{2}, \, 1 \right) \,,
\end{align}
motivated by two principles: that it is symmetric under the permutations which preserve the $s_{12}$ channel (i.e., swapping $p_1 \leftrightarrow p_2$), and that it contains few distinct prime factors.
The first condition reduces the number of permuted integral families we need to evaluate in order to obtain the boundary values.
The second condition reduces the number of independent transcendental constants appearing in the boundary values, which simplifies the construction of the basis of special functions. 
The order-$\eps^0$ boundary values $\vec{g}_{\tau}^{(0)}$ are rational constants. We obtain them up to their overall normalisation by solving the `first-entry conditions'~\cite{Gaiotto:2011dt}, i.e., by requiring the absence of unphysical branch cuts in the solutions. We fix the overall normalisation and the higher-order boundary values $\vec{g}_{\tau}^{(w)}(\vec{s}_0)$ (for $1\le w\le 4$) by evaluating all \acp{MI} with \texttt{AMFlow}~\cite{Liu:2022chg} (interfaced to 
\texttt{FiniteFlow}~\cite{Peraro:2019svx} and \texttt{LiteRed}~\cite{Lee:2012cn}) at $\vec{s}_0$ with at least $60$-digit precision.
We anticipate from \cref{sec:basis_MPLs} that, although we use floating-point boundary values, our results in terms of \acp{MPL} are fully~analytic. 

The canonical \acp{DE} \labelcref{eq:canonicalDEs} and the boundary values for all integral families are the input for the algorithm of \incite{Abreu:2023rco} for constructing a basis of special functions. We refer to the original work for a thorough discussion. 
Out of all \ac{MI} coefficients up to transcendental weight $4$, the algorithm selects a subset, denoted $F \coloneqq \{F^{(w)}_i(\vec{s})\}$, which satisfy two constraints. 
First, they are \emph{algebraically independent}, that is, there are no polynomial functional relations among them. Second, the \ac{MI} coefficients of all families (including all permutations of the external massless legs) up to transcendental weight $4$ are expressed as polynomials in the $\{F^{(w)}_i(\vec{s})\}$ and the zeta values $\zeta_2 = \pi^2/6$ and $\zeta_3$. For example, an arbitrary weight-$2$ \ac{MI} coefficient $g^{(2)}(\vec{s})$ has the general form
\begin{align}
g^{(2)}(\vec{s}) = \sum_{i=1}^{3} c_i \, F_i^{(2)}(\vec{s}) + \sum_{i \le j =1}^{4} d_{ij} \, F_i^{(1)}(\vec{s}) \,  F_j^{(1)}(\vec{s}) + e \, \zeta_2 \,,
\end{align}
with $c_i, d_{ij}, e \in \mathbb{Q}$. This special subset of \ac{MI} coefficients, $\{F^{(w)}_i(\vec{s})\}$, constitutes our special function basis. 
We give the number of functions in the basis in \cref{tab:func_basis}. Note that there is freedom in the choice of which \ac{MI} coefficients make up the basis. We make use of this freedom to choose as many basis-elements as possible from the one-loop family, then complement them with coefficients from the planar two-loop families, and finally complete them with coefficients from the non-planar two-loop families. In this way no two-loop \ac{MI} coefficients appear in the one-loop amplitudes, and no non-planar two-loop \ac{MI} coefficients appear in those amplitudes where only planar diagrams contribute (as is often the case in the leading colour approximation of QCD).

\begin{table}
    \centering
    \begin{tabular}{cc}
        \hline
        Weight & Number of basis functions \\
        \hline
        $1$ & 4 \\
        $2$ & 3 \\
        $3$ & 20 \\
        $4$ & 67 \\
        \hline
    \end{tabular}
    \caption{Number of functions $\{F^{(w)}_i\}$ in the basis weight by weight.}
    \label{tab:func_basis}
\end{table}

The folder \path{mi2func/} of our ancillary files~\cite{zenodo} contains the expression of all \ac{MI} coefficients (for all one- and two-loop integral families in all permutations of the external massless legs) up to weight $4$ in terms of our special function basis. This result enables the efficient amplitude-computation strategy based on finite-field arithmetic discussed in \cref{sec:calc}.
However, at this stage the basis functions $\{F^{(w)}_i\}$ are expressed in terms of Chen iterated integrals~\cite{Chen:1977oja} and numerical boundary values $\vec{g}^{(w)}(\vec{s}_0)$. This representation is excellent for investigating the analytic properties of Feynman integrals and amplitudes, but it is not readily suitable for an efficient numerical evaluation.
In the next section we discuss how we construct a representation of the function basis in terms of \acp{MPL} and zeta values, which is well suited for an efficient and stable numerical evaluation.

\subsection{Expression in terms of \aclp{MPL}}
\label{sec:basis_MPLs}

In this section we construct a representation of our function basis in terms of \acp{MPL}. 
Note that, while the basis functions $\{F^{(w)}_i\}$ are by construction algebraically independent,
the \acp{MPL} appearing in their representation constructed in this section may not be. Nonetheless,
we take a number of simple measures to reduce their number and optimise the representation.
The weight-$n$ \ac{MPL} of indices $\{a_1,\ldots,a_n\}$ and argument $x$ is defined recursively~as
\begin{align} 
G(a_1,a_2,\ldots,a_n; x) \coloneqq \int_0^x \frac{\dd t}{t - a_1} \, G(a_2,\ldots,a_n; t) \,, \qquad a_n \neq 0 \,, 
\end{align}
starting with $G(;x) = 1$. Trailing zeros, i.e., zeros in the right-most indices, are allowed through the definition
\begin{align} \label{eq:trailingzeros}
G(\underbrace{0,\ldots,0}_{k}; x) \coloneqq \frac{1}{k!} \log^k(x) \,.
\end{align}
We refer to \incite{Vollinga:2004sn} for a thorough discussion.

Since the letters in \cref{eq:alphabet} are rational and linear in all variables, we can solve the canonical \acp{DE} in \cref{eq:canonicalDEs} algorithmically in terms of \acp{MPL}. Order by order in $\eps$, the solution is given~by
\begin{align} \label{eq:solution}
\vec{g}^{(w)}_{\tau}(\vec{s}) = \sum_{i=1}^7 \mathrm{A}_i^{(\tau)} \cdot \int_{\gamma} \dd \log \bigl(W_i(\vec{s}=\gamma) \bigr) \, \vec{g}^{(w-1)}_{\tau}\bigl(\vec{s}=\gamma\bigr) + \vec{b}^{(w)}_{\tau} \,,
\end{align}
starting from the constant weight-$0$ boundary values $\vec{g}^{(0)}_{\tau}$ determined in the previous subsection. Here, $\gamma$ is a path connecting an arbitrary base-point $\vec{s}_{\mathrm{base}}$ to the end-point $\vec{s}$. The weight-$w$ constants $\vec{b}^{(w)}_{\tau} $ are given by the values of the integrals at the base-point, $ \vec{b}^{(w)}_{\tau} = \vec{g}^{(w)}_{\tau}(\vec{s}_{\mathrm{base}})$.
For $\vec{s}_{\mathrm{base}}$ we may use the boundary point $\vec{s}_0$ in \cref{eq:s0}, so that the constants $\vec{b}^{(w)}_{\tau}$ coincide with the boundary values determined numerically in the previous section. We follow a different approach, which allows us to trade all numerical constants in the expressions for zeta values.

We find it convenient to change variables from $(s_{12},s_{23},s_4)$ to $(z_1,z_2,s_4)$, with
\begin{align}
z_1 = \frac{s_{12}}{s_4} \,, \qquad \qquad z_2 = \frac{s_{23}}{s_4} \,.
\end{align}
This way, there is only one dimensionful variable, $s_4$, the dependence on which is fixed as an overall factor by dimensional analysis.
We then integrate the canonical \acp{DE} as in \cref{eq:solution} along the following piece-wise path in the $(z_1,z_2,s_4)$ space:
\begin{align} \label{eq:path}
(0,0,0) \overset{\gamma_1}{\longrightarrow} (z_1, 0, 0)  \overset{\gamma_2}{\longrightarrow} (z_1, z_2, 0)  \overset{\gamma_3}{\longrightarrow} (z_1, z_2, s_4) \,.
\end{align}
Since the Feynman integrals are divergent at the chosen base-point, the latter is understood in a \emph{regularised} sense (we refer to section~4 of \incite{Abreu:2022mfk} for a thorough discussion).
Choosing $(0,0,0)$ as base-point has the important benefit of removing spurious transcendental numbers that would pollute the solution were we to choose a base-point where the integrals are finite. As we will see below, only zeta values appear.
Roughly speaking, we define regularised, finite values $\vec{b}^{(w)}_{\tau} \coloneqq \mathrm{Reg} \, \vec{g}^{(w)}_{\tau}(\vec{s}_{\mathrm{base}})$ by introducing a regulator and formally setting to $0$ the (divergent) logarithms of the regulator.
Since the integrals are finite at a generic end-point $\vec{s}$, the divergences at the base-point must cancel out with divergences arising in the integration. We can thus drop all these divergences. Provided that we do it consistently between the integration and the base-point values $\vec{b}^{(w)}_{\tau}$, this leads to a finite and unique result. In practice, we fix the finite base-point values $\vec{b}^{(w)}_{\tau}$ by matching the solution $\vec{g}^{(w)}_{\tau}(\vec{s})$ evaluated at the boundary point $\vec{s}_0$ against the boundary values discussed in the previous subsection. 

We therefore keep the $\vec{b}^{(w)}_{\tau}$ as symbols and integrate the canonical \acp{DE} as in \cref{eq:solution} along the path in \cref{eq:path} up to weight $4$. 
We parameterise each piece of the path in \cref{eq:path} linearly. For instance, 
$\gamma_2(t) = ( z_1,  t  , 0 )$,
with $t \in [0,z_2]$.
\begin{itemize}
\item The $\gamma_1$ integration leads to \acp{MPL} with indices in $\{0,1\}$ and argument~$z_1$.
\item The $\gamma_2$ integration leads to \acp{MPL} with indices in $\{0,1, 1-z_1, -z_1\}$ and argument~$z_2$.
\item The $\gamma_3$ integration leads to powers of $\log(-s_4)$, fixed by dimensional analysis.
\end{itemize}
Once we have obtained expressions for all \acp{MI} in terms of \acp{MPL} and symbolic constants $\vec{b}^{(w)}_{\tau}$, we equate them to the numerical boundary values at $\vec{s}_0$, and solve for the $\vec{b}^{(w)}_{\tau}$. We use \texttt{GiNaC}~\cite{Bauer:2000cp,Vollinga:2004sn} to evaluate the \acp{MPL} numerically.
Finally, we use the \texttt{PSLQ} algorithm~\cite{PSLQ} to express the ensuing values of $\vec{b}^{(w)}_{\tau}$ in terms of $\zeta_2$ and $\zeta_3$. As a result, we obtain a fully analytic representation of all \acp{MI} --- and thus of our special function basis $\{F^{(w)}_i\}$ --- in terms of \acp{MPL} and zeta values, up to weight $4$.

Contrary to the functions in the basis $\{F^{(w)}_i\}$, the \acp{MPL} in their representation satisfy functional relations. We make use of this freedom to optimise our expressions in view of their numerical evaluation by reducing the number of distinct \acp{MPL} that need to be evaluated. First, we use the shuffle algebra of \acp{MPL} to push all trailing zeros into logarithms through \cref{eq:trailingzeros}~\cite{Vollinga:2004sn}. Next, we employ the scaling relation
\begin{align}
G(a_1, \ldots, a_n; x) = G\left(\frac{a_1}{x}, \ldots, \frac{a_n}{x}; 1\right) \,,
\end{align}
which holds for $x, a_n \neq 0$. As a result, all \acp{MPL} have argument $1$ and indices
\begin{align} \label{eq:indices}
l_0 = 0 \,, \qquad 
l_1 = \frac{s_4}{s_{12}} \,, \qquad 
l_2 = \frac{s_4}{s_{23}} \,, \qquad
l_3 = \frac{s_4-s_{12}}{s_{23}} \,, \qquad
l_4 = - \frac{s_{12}}{s_{23}}\,.
\end{align}
Finally, we decompose the \acp{MPL} to \emph{Lyndon words}~\cite{Radford1979ANR} using \texttt{PolyLogTools}~\cite{Duhr:2019tlz}; we refer to the latter work for a thorough explanation, and give here only a simple example. 
This procedure requires that we choose a symbolic ordering of the \ac{MPL} indices. We choose $l_0 \prec l_1 \prec l_2 \prec l_3 \prec l_4$, meaning that $l_1$ is greater than $l_0$, and so on. 
Consider the \ac{MPL} $G(l_1, l_0; 1)$, whose indices are not sorted according to the ordering above, since $l_1 \succ l_0$. We can use the shuffle algebra of \acp{MPL} to rewrite it in terms of \acp{MPL} whose indices are sorted according to the chosen ordering, as
\begin{align}
G(l_1, l_0; 1) = G(l_0;1) \, G(l_1;1) - G(l_0, l_1; 1) \,.
\end{align}
Doing this consistently throughout all expressions reduces the number of higher-weight \acp{MPL} in favour of products of lower-weight ones, which are cheaper to evaluate numerically. To maximise the impact in this sense, we tested all possible orderings of the indices and selected the one --- given above --- which minimises the number of weight-4 \acp{MPL}.
The resulting representation of the function basis contains $4$ weight-1, 6 weight-2, 19 weight-3, and 25 weight-4 \acp{MPL}, as well as $3$ logarithms: 
\begin{align} \label{eq:logs}
\log(s_{12}/s_4) \,, \qquad \quad \log(s_{23}/s_4) \,, \qquad \quad \log(-s_4) \,.
\end{align}
We write the latter in terms of logarithms rather than \acp{MPL} as they play an important role in the factorisation of the \ac{IR} divergences in the scattering amplitudes (see \cref{app:poles} for the \ac{IR} structure of the amplitudes we compute here). We stress that $\log(-s_4)$ is the only function of a dimensionful argument in our representation of the function basis. 

We provide in the folder \path{mi2func/} of our ancillary files~\cite{zenodo} the expression of the basis functions $\{F^{(w)}_i\}$ in terms of \acp{MPL}, logarithms, $\zeta_2$ and $\zeta_3$.

\smallskip

It is important to stress that the \acp{MPL} are multi-valued functions.
For unit argument, there is a pole on the integration contour whenever one of the indices lies between $0$ and $1$. In this case the contour must be deformed in the complex plane, either above or below the pole, leading to different branches. Our \acp{MPL} are thus well-defined only in the 
kinematic region where all \ac{MPL} indices in \cref{eq:indices} are either less than 0 or greater than 1, and we need $s_4 < 0$ for the argument of all logarithms in \cref{eq:logs} to be positive. 
We analytically continue the \acp{MPL} to the kinematic regions of interest by adding infinitesimal imaginary parts to the indices in the numerical evaluation, as we discuss in \cref{app:an_cont}.
The analytic continuation of these functions has been analysed in great detail in \incite{Gehrmann:2002zr}, 
where it is achieved by suitable changes of variables so that the imaginary parts are extracted explicitly, and all \acp{MPL} are real and single valued. 
This approach could lead to faster evaluations although it was unnecessary for the applications considered here.

\subsection{Performance and validation}
\label{sec:performance}

We validated our results for the \acp{MI} of all families by crosschecking them against values obtained with \texttt{AMFlow}~\cite{Liu:2022chg} at several random points in all the physical kinematic regions discussed in \cref{app:an_cont}. Furthermore, we find agreement with the results of~\incite{Gehrmann:2023etk}. We employ \texttt{GiNaC}~\cite{Bauer:2000cp,Vollinga:2004sn} to evaluate the \acp{MPL}.

Our results allow for an efficient and stable evaluation of the \acp{MI}, and are thus ready for immediate deployment in phenomenology. Indeed, the amplitudes we computed in this work have already been implemented in \texttt{McMule}~\cite{Banerjee:2020rww,ulrich_yannick_2022_6046769} to provide the \acl{RVV} electron-line corrections to $e \mu \to e \mu$ scattering. The evaluation is efficient, running at $\approx 130$ events per second in the bulk of the phase space~\cite{ulrich-radcor} using \texttt{handyG}~\cite{Naterop:2019xaf} for the evaluation of the \acp{MPL}.

\section{Conclusions}
\label{sec:conc}

In this article, we calculated analytically the two-loop \ac{QED} helicity amplitudes for the process $0\to\ell\bar\ell\gamma\gamma^*$  in
terms of a basis of \aclp{MPL} that are suitable for fast and
stable numerical evaluation. We employed modern finite-field evaluation
techniques to reconstruct the amplitudes directly in terms of the special
function basis, sidestepping the symbolic computation in all intermediate stages. As a by-product we have recomputed all two-loop master
integrals for four-point functions with an off-shell leg up to transcendental weight four, and provide all the
necessary ingredients needed to use them in amplitude computations with the
same kinematics.

We hope these new results will now open the path to \ac{N3LO} predictions that
can be used for the future MUonE experiment.

\medskip

\textit{Note added:} Another computation of these amplitudes~\cite{Fadin:2023phc} became available on the arXiv shortly after ours.

\acknowledgments

We thank Yannick Ulrich for providing the one-loop crosschecks, correspondence on the \texttt{McMule} implementation of these amplitudes, and other useful discussions. We further thank Heribertus Bayu Hartanto and Tiziano Peraro for collaboration on the codebase.
SZ wishes to thank Dmitry Chicherin and Vasily Sotnikov for many useful discussions.
This project received funding from the European Union's Horizon 2020 research and innovation programme \textit{High precision multi-jet dynamics at the LHC} (grant agreement number 772099).

\appendix

\section{Definition of the Feynman integral families}
\label{app:int_def}

For each two-loop integral family $\tau$ corresponding to one of the maximal topologies shown in \cref{fig:int-fams},
the Feynman integrals have the form
\begin{equation}
	j^{\tau}(a_1, \ldots, a_9) = \mathrm{e}^{2 \eps \gamma_{E}} \int \frac{\mathrm{d}^{4-2\eps} k_1}{\mathrm{i} \pi^{2-\eps}} \frac{\mathrm{d}^{4-2\eps} k_2}{\mathrm{i} \pi^{2-\eps}} \frac{1}{D_{\tau,1}^{a_1} \ldots D_{\tau,9}^{a_9}} \,.
\end{equation}
The sets $\{D_{\tau,1}, \ldots, D_{\tau,9}\}$ contain seven (inverse) propagators and two \acp{ISP} ($a_8, a_9 \le 0$). 
For the maximal topologies under consideration, they are given by\footnote{We use a naming convention analogous to that of \incite{Abreu:2020jxa}.}:
\begin{itemize}
	\item penta-triangle, \textbf{mzz} configuration:
		\begin{align} \begin{aligned}
			\big\{k_1^2,(k_1+p_1+p_2+p_3)^2,(k_1+p_2+p_3)^2,(k_1+p_3)^2,k_2^2,(k_2-p_3)^2,\\(k_1+k_2)^2,(k_2-p_1-p_2-p_3)^2,(k_2-p_2-p_3)^2\big\} \,,
		\end{aligned} \end{align}
	\item penta-triangle, \textbf{zmz} configuration:
		\begin{align} \begin{aligned}
		\big\{k_1^2,(k_1-p_1)^2,(k_1+p_2+p_3)^2,(k_1+p_3)^2,k_2^2,(k_2-p_3)^2,(k_1+k_2)^2,\\(k_2+p_1)^2,(k_2-p_2-p_3)^2 \big\} \,,
		\end{aligned} \end{align}
	\item penta-triangle, \textbf{zzz} configuration:
		\begin{align} \begin{aligned}
			\big\{k_1^2,(k_1-p_1)^2,(k_1-p_1-p_2)^2,(k_1-p_1-p_2-p_3)^2,k_2^2,(k_2+p_1+p_2+p_3)^2, \\ (k_1+k_2)^2,(k_2+p_1)^2,(k_2+p_1+p_2)^2\big\} \,,
		\end{aligned} \end{align}
	\item planar double-box:
		\begin{align} \begin{aligned}
			\big\{k_1^2,(k_1-p_1)^2,(k_1-p_1-p_2)^2,k_2^2,(k_2+p_1+p_2+p_3)^2,(k_2+p_1+p_2)^2, \\ (k_1+k_2)^2,(k_1-p_1-p_2-p_3)^2,(k_2+p_1)^2\big\} \,,
		\end{aligned} \end{align}
	\item crossed double-box, \textbf{mz} configuration:
		\begin{align} \begin{aligned}
			\big\{k_1^2,(k_1+p_1+p_2+p_3)^2,(k_1+p_2+p_3)^2,k_2^2,(k_2-p_2)^2,(k_1+k_2)^2,\\ (k_1+k_2+p_3)^2,(k_1+p_3)^2,(k_2-p_1-p_2-p_3)^2\big\} \,,
		\end{aligned} \end{align}
	\item crossed double-box, \textbf{zz} configuration:
		\begin{align} \begin{aligned}
			\big\{k_1^2,(k_1-p_1)^2,(k_1-p_1-p_2)^2,k_2^2,(k_2-p_3)^2,(k_1+k_2)^2, \\ (k_1+k_2-p_1-p_2-p_3)^2,(k_1-p_1-p_2-p_3)^2,(k_2+p_1)^2\big\} \,.
		\end{aligned} \end{align}
\end{itemize}
We also use the one-loop (one-mass) box family, made of the following integrals:
\begin{equation}
	j^{\rm box}(a_1, a_2, a_3, a_4) = \mathrm{e}^{\eps \gamma_{E}} \int \frac{\mathrm{d}^{4-2\eps} k}{\mathrm{i} \pi^{2-\eps}} \frac{1}{D_{\mathrm{box}, 1}^{a_1} D_{\mathrm{box},2}^{a_2} D_{\mathrm{box},3}^{a_3} D_{\mathrm{box},4}^{a_4}} \,,
\end{equation}
with the four inverse propagators $D_{\mathrm{box},i}$
	\begin{equation}
		\big\{k_1^2, (k_1-p_1)^2, (k_1-p_1-p_2)^2, (k_1-p_1-p_2-p_3)^2 \big\} \,.
	\end{equation}
Feynman's prescription for the imaginary parts of all propagators is implicit.

These family definitions (strictly with the ordering of inverse propagators and \acp{ISP} shown above) correspond to the integrals \texttt{j[family,$a_1,$\ldots]} that build the canonical \ac{MI} bases provided in the \path{pure_mi_bases/} directory of our ancillary files~\cite{zenodo}. In this notation, each \texttt{j[...]} represents a Feynman integral within a given integral family, while the numbers $a_i$ refer to the powers of its propagators and \acp{ISP}.

\section{Optimised \acs{IBP} reduction procedure for amplitudes with many permuted integral families}
\label{app:altIBPs}
An amplitude will in general have contributions from permutations of
the \textit{ordered} integral families shown in figure \ref{fig:int-fams}. To reduce the tensor integrals in the amplitude, \ac{IBP} identities must
be generated for all the permutations of these ordered families. This can
lead to a very large \ac{IBP} system. 
The performance of the reduction setup is extremely sensitive to the number of \ac{IBP}
identities required so, to minimise the memory consumption, we choose to
generate \ac{IBP} identities only for the ordered families. Next, we obtain the reduction for any permutation of these families by permuting the `ordered' reduction numerically over finite fields.
The result is then given in terms of \acp{MI} of each family permutation, but it is missing the symmetry relations that can be found between subsectors of different families. To express the final result in terms of a minimal set of \acp{MI}, we find such relations
from a separate computation. One may account for integral symmetries using automated tools such as \texttt{LiteRed}~\cite{Lee:2012cn}. Since we use a pure basis of \acp{MI}, the symmetry relations
amongst them will have rational numbers as coefficients. This is because the presence of any kinematic invariant would spoil the purity of the canonical \acp{DE} (see \cref{sec:spec-fns}), and would mean that such a symmetry relation in fact involves non-pure integrals. Therefore, the computation of the missing symmetry relations can be performed with all kinematic invariants set to numeric values, which significantly lowers the complexity of this task. Finally, we note that even if symmetries amongst the
\acp{MI} were missed, a representation of the integrals in terms of a basis of
special functions ---~as we construct in \cref{sec:spec-fns}~--- would automatically incorporate the extra simplifications and
so the same final result would be obtained. Nonetheless, in practice we do find it useful to include these symmetry relations, as they reduce the number of independent coefficients that have to be processed further.

The procedure can be summarised as follows:
\begin{enumerate}
	\item Generate (analytic) IBPs for the six ordered families.
	\item Compute the mappings between permutations of the \acp{MI} of the system above.
	\item Take the tensor integrals in the amplitudes for each permutation of these families and solve the linear system over finite fields.
	\item Apply the symmetry mappings between the \acp{MI} of each family permutation to find the minimal set for the full system.
\end{enumerate}

Since there are a few additional bits of terminology, we can consider a concrete
example to clarify everything. At one-loop, a four-point process with a
single off-shell leg can be described by a single independent integral family
which is simply the box topology (see \cref{app:int_def} for its explicit definition). Following the Laporta reduction algorithm leads to a basis
of four \acp{MI},
\begin{equation}
	{\rm MI}^{\rm box} = \{j^{\rm box}(1,1,1,1),\, j^{\rm box}(1,0,1,0),\, j^{\rm box}(0,1,0,1),\, j^{\rm box}(1,0,0,1)\} \,,
\end{equation}
which are the scalar box and scalar bubble integrals in channels $s_{12},
s_{23}$ and $s_4$ respectively. An amplitude will, in general, be written in
terms of three permutations of this family. Let us denote
these permutations as $j^{\rm box, 1234}$, $j^{\rm box, 2314}$, and $j^{\rm box, 3124}$,
where $j^{\rm box, 1234} = j^{\rm box}$ as above and the additional superscript indices refer to the order of the external legs. Following our procedure we
would load one set of IBP relations generated for $j^{\rm box}$. These
identities can then be permuted numerically, for example as \texttt{FiniteFlow} graphs, to reduce
tensor integrals in each of the three permuted families. The result is now in terms
of twelve \acp{MI}: three boxes and nine bubbles. While the amplitude is already
in a minimal basis of box integrals, there is clearly an over-complete set of
bubbles. The independent bubbles are in the channels $s_{12}$, $s_{23}$,
$s_{13}$, and $s_4$, so the five additional symmetry mappings are
\begin{align}
	\begin{aligned}
		j^{\rm box, 2314}(1,0,1,0) &= j^{\rm box, 1234}(0,1,0,1) \,, &
		j^{\rm box, 3124}(1,0,1,0) &= j^{\rm box, 2314}(0,1,0,1) \,, \\
		j^{\rm box, 3124}(0,1,0,1) &= j^{\rm box, 1234}(1,0,1,0) \,, &
		j^{\rm box, 2314}(1,0,0,1) &= j^{\rm box, 1234}(1,0,0,1) \,, \\
		j^{\rm box, 3124}(1,0,0,1) &= j^{\rm box, 1234}(1,0,0,1) \,.
	\end{aligned}
\end{align}
After applying these identities we arrive at the final result with seven
\acp{MI} which cover all permutations of the integral families. 
This approach would not lead to any significant performance
enhancements in this simple example of course,
but it can be particularly important when considering
high-multiplicity examples where the number of permutations is
high.

\section{Rational parametrisation of the kinematics}
\label{app:mtvs}

Since we are applying finite-field techniques to helicity amplitudes, we employ a rational parametrisation of the external kinematics using Hodges's momentum twistor formalism~\cite{Hodges:2009hk}.
While this is not essential to combat the algebraic complexity for the kinematics considered here, it does provide a convenient parametrisation of the spinor products.

The single-off-shell four-particle phase space $p$ is obtained from a massless five-particle parametrisation $q$ (defined in appendix A of \incite{Badger:2021imn} with $\{x_2\leftrightarrow x_4,x_3\leftrightarrow x_5\}$) under
\begin{align}
    p_i &= q_i \qquad \forall i=1,2,3, & p_4=q_4+q_5 \, .
\end{align}
The momentum twistor variables $x_i$ for $p$ are then related to the scalar invariants $\vec{s}$ through
\begin{align} \label{eq:mtvs}
    s_{12} &= x_1 \,, &
    s_{23} &= x_1 x_2 \,, &
    s_4 &= x_1 x_3 \,.
\end{align}

Momentum twistors allow us to express any spinor expression as a rational function in the variables $x_i$.
In this representation the helicity scaling is however obscured, as we have fixed the spinor phases in order to achieve a parameterisation in terms of the minimal number of variables (see e.g.~\incite{Badger:2016uuq}).
Therefore, we need to manually restore the phase information at the end of the computation.
This can be achieved by multiplying the momentum twistor expression by an arbitrary factor $\Phi$ with the same helicity scaling as the helicity amplitude under consideration, divided by that factor written in terms of momentum twistor variables.
For example, for the helicity configurations of \cref{eq:helconfs}, we can use the phase factors
\begin{align}
  \Phi(-++) &= \frac{\langle 1 2 \rangle}{\langle 2 3 \rangle^2} \,,&
  \Phi(-+-) &= \frac{[ 1 2 ]}{[ 1 3 ]^2} \,,
\end{align}
which in our momentum twistor parameterisation are given by
\begin{align}
  \Phi(-++) &= x_1^2 \,,&
  \Phi(-+-) &= - \frac{1}{x_1 (1 + x_2 - x_3)^2}   \,.
\end{align}
We refer to appendix~C of \incite{Badger:2023mgf} for a thorough discussion of how to restore the phase information in a momentum twistor parameterisation.

\section{Renormalisation and \acl{IR} structure}
\label{app:poles}

We renormalise the coupling constant by trading the bare coupling $\alpha_{\mathrm{bare}}$ for the renormalised one $\alpha_{\mathrm{R}}$ through
\begin{align}
    \label{eq:alpha-bare}
    \alpha_{\mathrm{bare}} = \alpha_{\mathrm{R}}(\mu_R) \, Z_{\alpha}\bigl(\alpha_{\mathrm{R}}(\mu_R) \bigr)  \, \mu_R^{2 \eps} \, S_{\eps} \,,
\end{align}
with $S_\eps=(4\pi)^{-\eps} \mathrm{e}^{\eps\gamma_E}$.
The renormalisation factor $Z_{\alpha}$ in the $\overline{\text{MS}}$ scheme is \cite{Barnreuther:2013qvf,Bonciani:2021okt}
\begin{align}
    \label{eq:Za}
    Z_{\alpha}(\alpha) = 1 - \frac{\alpha}{4 \pi} \frac{\beta_0}{\eps} -
    \left(\frac{\alpha}{4 \pi}\right)^2 \left( -\frac{\beta_0^2}{\eps^2} + \frac{1}{2}\frac{\beta_1}{\eps} \right) + \mathcal{O}\bigl(\alpha^3\bigr) \,.
\end{align}
The $\beta$-function is defined from the renormalised coupling as
\begin{align} \label{eq:beta_func_def}
\frac{\dd \alpha_{\rm R}(\mu_R)}{\dd \ln \mu_R} = \left[ -2 \, \eps + \beta\bigl(\alpha_{\rm R}(\mu_R)\bigr) \right]  \alpha_{\rm R}(\mu_R) \,,
\end{align}
and expanded as
\begin{align} \label{eq:beta_func}
\beta( \alpha ) = -2 \, \frac{\alpha}{4 \pi}  \sum_{k\ge 0} \beta_k \left(\frac{\alpha}{4\pi} \right)^k \,,
\end{align}
with 
\begin{align} \label{eq:beta_coeffs}
\beta_0 = -\frac{4}{3} \nl \,, \qquad \qquad \beta_1 = - 4 \, \nl \,.
\end{align}
The photon wavefunction renormalisation factor is $Z_A=Z_\alpha$, which we include due to the external off-shell photon.
The complete renormalisation procedure then is
\begin{align}
    \mathcal{A}^\mu_\text{renorm}(\alpha_R) = Z_A^{\frac{1}{2}}(\alpha_R) \, \mathcal{A}^\mu_\text{bare}(\alpha_\text{bare}) \, ,
\end{align}
where $\alpha_\text{bare}$ is expressed in terms of $\alpha_R$ through \cref{eq:alpha-bare}.

The \ac{IR} poles of the renormalised amplitude factorise as~\cite{Catani:1998bh,Gardi:2009qi,Gardi:2009zv,Becher:2009cu,Becher:2009qa}
\begin{align}
 \mathcal{A}^\mu_\text{renorm}(\alpha_R) = Z(\alpha_R) \,  \mathcal{F}^\mu(\alpha_R) \,,
\end{align}
so that $ Z(\alpha_R)$ captures all \ac{IR} poles and $\mathcal{F}^\mu$ is a finite remainder.
We obtain the explicit two-loop expression of the \ac{IR} factor $Z(\alpha_R)$ by
choosing \ac{QED} parameters ($C_A=0$, $C_F=1$, and $T_F=1$) in the non-abelian gauge-theory expressions of \incite{Becher:2009qa}. We expand it as
\begin{align}
    Z(\alpha) &= \sum_{k\ge0} Z^{(L)} \left(\frac{\alpha}{4\pi}\right)^L \,.
\end{align}
The coefficients $Z^{(L)}$ are expressed in terms of the anomalous dimension
\begin{align}
    \Gamma &= \gamma^\text{cusp}\ln\left(\frac{-s_{12}}{\mu^2}\right)+2\gamma^l+\gamma^A \,,
\end{align}
and its derivative
\begin{align}
    \Gamma^\prime &\coloneqq \frac{\partial\Gamma}{\partial\ln\mu} = -2\gamma^\text{cusp} \,.
\end{align}
Here, $\gamma^\text{cusp}$ is the cusp anomalous dimension, while $\gamma^l$ and $\gamma^A$ are the lepton's and the photon's collinear anomalous dimensions, respectively.
We expand all anomalous dimensions $y\in\{\Gamma,\gamma^i\}$ as
\begin{align}
    y &= \frac{\alpha}{4\pi} \sum_{k\ge0} y_k \left(\frac{\alpha}{4\pi}\right)^k \,,
\end{align}
with coefficients
\begin{subequations}
\begin{align}
    \gamma^l_0 &= -3 \,, &
    \gamma^l_1 &= -\frac{3}{2}+2\pi^2-24 \, \zeta_3+\nl\left(\frac{130}{27}+\frac{2}{3}\pi^2\right) \,, \\
    \gamma^A_0 &= -\beta_0 \,, &
    \gamma^A_1 &= -\beta_1 \,, \\
    \gamma^\text{cusp}_0 &= 4 \,, &
    \gamma^\text{cusp}_1 &= -\frac{80}{9} \nl \,.
\end{align}
\end{subequations}
Finally, the coefficients of the \ac{IR} factor $Z$ up to two loop are given by
\begin{align}
    \label{eq:ir-pole-coeffs}
    Z^{(0)} &= 1 \,, &
    Z^{(1)} &= \frac{\Gamma_0^\prime}{4\eps^2}+\frac{\Gamma_0}{2\eps} \,, &
    Z^{(2)} &= \frac{{Z^{(1)}}^2}{2} -\frac{3\beta_0\Gamma_0^\prime}{16\eps^3}+\frac{\Gamma_1^\prime-4\beta_0\Gamma_0}{16\eps^2}+\frac{\Gamma_1}{4\eps} \,.
\end{align}

Putting together the subtraction of \ac{UV} and \ac{IR} poles, and expanding the resulting finite remainder $\mathcal{F}^{\mu}(\alpha_R)$ in $\alpha_R$ leads to the definitions in \cref{eq:finite-remainders}.

\section{Analytic continuation}
\label{app:an_cont}

We analytically continue the \acp{MPL} by adding an infinitesimal positive (or negative) imaginary part to the \ac{MPL} indices $l_i$ in \cref{eq:indices} whenever they fall between $0$ and $1$. The imaginary part of each index prescribes how to deform the integration contour around the pole associated with it. We do similarly for the logarithms in \cref{eq:logs}.
To this end, following \incite{Gehrmann:2002zr}, we change variables from $(s_{12},s_{23},s_4)$ to $(s_{12},s_{23},s_{13})$, with $s_4 = s_{12} + s_{23} + s_{13}$. We then add a small positive imaginary part to the latter variables, as
\begin{align} \label{eq:add_im}
s_{12} \longrightarrow s_{12} + \mathrm{i} \, c_{1} \, \delta \, , \qquad
s_{23} \longrightarrow s_{23} + \mathrm{i} \, c_{2} \, \delta  \, , \qquad
s_{13} \longrightarrow s_{13} + \mathrm{i} \, c_{3} \, \delta  \, , 
\end{align}
where $c_{1}$, $c_{2}$ and $c_{3}$ are arbitrary positive constants, and $\delta$ is a positive infinitesimal. 
Finally, we check whether this substitution gives a positive or negative imaginary part to each \ac{MPL} index $l_i$.
This depends on the domain of the kinematic variables.
We focus on three kinematic regions which are of phenomenological interest. The analytic continuation for any other region may be obtained similarly.

\smallskip

\paragraph{Electron-line corrections to $e^- \mu^- \to e^- \mu^- \gamma$.}
To define the domain of the kinematic variables relevant for this application, 
we embed the four-particle off-shell process of \cref{eq:scatter} in the five-particle process $e^- \mu^- \to e^- \mu^- \gamma$. We then determine the kinematic constraints for the five-particle process (see e.g.\ appendix~A of \incite{Chicherin:2021dyp}), and from them derive the constraints on the four-point off-shell kinematics. The result is
\begin{align}
\label{eq:region_emu-emugamma}
\mathcal{P}_{e\mu\to{e}\mu\gamma} \coloneqq \{\vec{s} \colon s_{12} < 0 \, \land \, s_{23} < 0 \, \land \, 0 < s_{13} < -s_{12} - s_{23} \} \,.
\end{align}
The \ac{MPL} index $l_4 = - s_{12}/s_{23}$ is always negative in $\mathcal{P}_{{e}\mu\to{e}\mu\gamma}$, hence no analytic continuation is required. The other three indices may instead fall between $0$ and $1$. Let us study $l_1$. Changing variables from $s_4$ to $s_{13}$ and adding imaginary parts as in \cref{eq:add_im} gives
\begin{align}
l_1 = \frac{s_{12} + s_{13} + s_{23}}{s_{12}} +  \frac{\mathrm{i} \delta}{s_{12}^2} \left[ (c_2 + c_3) s_{12} - c_1 (s_{13} + s_{23}) \right] + \mathcal{O}\left(\delta^2\right) \,.
\end{align}
The imaginary part of $l_1$ may be either negative or positive in $\mathcal{P}_{{e}\mu\to{e}\mu\gamma} $. However, it is strictly negative in the subregion of $\mathcal{P}_{{e}\mu\to{e}\mu\gamma} $ where $0<l_1<1$. We therefore assign a negative imaginary part to $l_1$ whenever $0<l_1<1$ in $\mathcal{P}_{{e}\mu\to{e}\mu\gamma} $. The analysis of the other indices follows similarly, and is summarised in \cref{tab:an_cont}.
The arguments of the three logarithms in \cref{eq:logs} are positive in $\mathcal{P}_{e\mu\to{e}\mu\gamma}$.

\begin{table}
    \centering
    \begin{tabular}{cccc}
        \hline
        Index & $\mathcal{P}_{{e}\mu\to{e}\mu\gamma}$ & $\mathcal{P}_{{e}\bar{{e}}\to\gamma \gamma^*}$ & $\mathcal{P}_{\gamma^*\to e \bar{e} \gamma}$  \\
        \hline
        $l_1$  & $-$ & $+$ & $0$ \\
        $l_2$  & $-$ & $0$ & $0$ \\
        $l_3$  & $-$ & $0$ & $0$ \\
        $l_4$  & $0$ & $0$ & $0$ \\
        \hline
    \end{tabular}
    \caption{Imaginary parts of the \ac{MPL} indices defined by \cref{eq:indices} in the three kinematic regions discussed in \cref{app:an_cont}. The symbol $+$ ($-$) denotes a positive (negative) imaginary part, while $0$ means no analytic continuation is needed.}
    \label{tab:an_cont}
\end{table}

\smallskip

\paragraph{Corrections to $e^- e^+ \to \gamma \gamma^*$.} The relevant domain of the kinematic variables in this case can be derived directly for the four-point kinematics, and is typically named the $s_{12}$ channel. It is given by 
\begin{align}
\label{eq:region_eebar}
\mathcal{P}_{e\bar{e}\to\gamma\gamma^*} \coloneqq \{\vec{s} \colon s_{23} < 0 \, \land \, s_{13} < 0 \, \land \, s_{12} > - s_{23} - s_{13} \} \,.
\end{align}
The \ac{MPL} indices $l_2$, $l_3$ and $l_4$ can never fall between $0$ and $1$ in $\mathcal{P}_{e\bar{e}\to\gamma\gamma^*}$, and hence require no analytic continuation. We instead need to add a positive imaginary part to $l_1$. 
In this region also the logarithms in \cref{eq:logs} need to be analytically continued. The argument of $\log(s_{12}/s_4)$ is positive in $\mathcal{P}_{e\bar{e}\to\gamma\gamma^*}$. By adding imaginary parts to the arguments of the other logarithms and studying them where the arguments are negative in $\mathcal{P}_{e\bar{e}\to\gamma\gamma^*}$, we determine that the analytic continuation is achieved through the following replacements:
\begin{align}
\log(s_{23}/s_4) \longrightarrow \log(-s_{23}/s_{4})+ \mathrm{i} \pi \,, \qquad  \log(-s_4) \longrightarrow \log(s_4)- \mathrm{i} \pi \,.
\end{align}

\paragraph{Corrections to the decay $\gamma^* \to e^- e^+ \gamma$.} The relevant domain of the kinematic variables~is
\begin{align}
\label{eq:region_decay}
\mathcal{P}_{\gamma^*\to e \bar{e} \gamma} \coloneqq  \{ \vec{s} \colon s_{12} > 0 \, \land \, s_{23} > 0 \, \land \,  s_{13} > 0 \} \,.
\end{align}
All \ac{MPL} indices $l_i$ in \cref{eq:indices} are either $l_i < 0$ or $l_i > 1$, hence no analytic continuation is required. The same holds for the first two logarithms in \cref{eq:logs}, whose arguments are positive. The only function which needs to be analytically continued is $\log(-s_4)$. We achieve this by replacing
\begin{align}
\log(-s_4) \longrightarrow \log(s_{4}) - \mathrm{i} \pi \,.
\end{align}

\smallskip

The information about the imaginary parts of the \ac{MPL} indices can be fed into the publicly available libraries for evaluating these functions numerically, such as \texttt{FastGPL}~\cite{Wang:2021imw}, \texttt{GiNaC}~\cite{Bauer:2000cp,Vollinga:2004sn}, and \texttt{handyG}~\cite{Naterop:2019xaf}. This typically leads to longer evaluation times with respect to \acp{MPL} which do not need analytic continuation. We find that this is not an issue for the planned applications of our results (see \cref{sec:performance}). Nonetheless, we note that a more performant evaluation may be achieved by tailoring the representation to the kinematic region of interest in such a way that no \acp{MPL} require analytic continuation. We refer to \incite{gehrmann:2000zt,gehrmann:2001ck,Gehrmann:2002zr,Duhr:2012fh,Gehrmann:2023etk} for a detailed discussion.

\bibliographystyle{JHEP}
\bibliography{bibliography}

\end{document}